\documentclass[twoside, 12pt, a4paper]{article}

\usepackage{amsmath}
\usepackage{amssymb}
\usepackage{epsfig}
\usepackage{lscape}
\usepackage{url}
\usepackage{xcolor}
\usepackage{graphicx}
\usepackage{subfig}

\newcommand{\half}{{\textstyle \frac{1}{2}}}
\newcommand{\kam}{{\sc kam}}
\newcommand{\FPU}{{\sc fpu}}

\newcommand{\opin}[2]{\mathopen] #1, #2 \mathclose[}
\newcommand{\clopin}[2]{\mathopen[ #1, #2 \mathclose[}
\newcommand{\opclin}[2]{\mathopen] #1, #2 \mathclose]}

\newcommand{\X}[1]{X_{\textstyle \! \mbox{\small $#1$}}}

\newcommand{\Sum}{{\displaystyle \sum}}

\newcommand{\re}{{\mathop{\rm Re}}\,}
\newcommand{\im}{{\mathop{\rm Im}}\,}

\newcommand{\diag}{{\mathop{\rm diag}}}

\newcommand{\D}{{\rm d}}
\newcommand{\E}{{\rm e}}
\newcommand{\I}{{\rm i}}

\newcommand{\R}{\mathbb{R}}

\newcommand{\Z}{\mathbb{Z}}

\newcommand{\pS}{\mathbb{S}^1}

\newcommand{\cH}{\mathcal H}
\newcommand{\cK}{\mathcal K}

\newtheorem{theorem}{Theorem}[section]
\newtheorem{lemma}[theorem]{Lemma}
\newtheorem{proposition}[theorem]{Proposition}
\newtheorem{corollary}[theorem]{Corollary}
\newtheorem{definition}[theorem]{Definition}
\newtheorem{remark}[theorem]{Remark}
\newtheorem{example}[theorem]{Example}

\newenvironment{Remark}{\begin{remark}
       \begin{rm}}{\end{rm} \end{remark}}

\begin{document}

\title{\protect\Large The $1{:}2{:}4$ resonance in a particle chain}

\author{{\protect\normalsize Heinz Han{\ss}mann} \protect\\[-1mm]
   {\protect\footnotesize\protect\it Mathematisch Instituut,
           Universiteit Utrecht} \protect\\[-2mm]
   {\protect\footnotesize\protect\it Postbus 80010,
           3508~TA Utrecht, The Netherlands} \protect\\
   {\protect\normalsize Reza Mazrooei-Sebdani} \protect\\[-1mm]
   {\protect\footnotesize\protect\it Department of Mathematical Sciences,
           Isfahan University of Technology} \protect\\[-2mm]
   {\protect\footnotesize\protect\it Isfahan 84156-83111,
           Iran} \protect\\
   {\protect\normalsize Ferdinand Verhulst} \protect\\[-1mm]
   {\protect\footnotesize\protect\it Mathematisch Instituut,
           Universiteit Utrecht} \protect\\[-2mm]
   {\protect\footnotesize\protect\it Postbus 80010,
           3508~TA Utrecht, The Netherlands}}

\date{\protect\normalsize 27 January 2020}

\maketitle

\begin{abstract}
\noindent
We consider four masses in a circular configuration with
nearest-neighbour interaction, generalizing the spatially
periodic Fermi--Pasta--Ulam-chain where all masses are equal.
We identify the mass ratios that produce the $1{:}2{:}4$~resonance
--- the normal form in general is non-integrable already at cubic
order.
Taking two of the four masses equal allows to retain a discrete
symmetry of the fully symmetric Fermi--Pasta--Ulam-chain and yields
an integrable normal form approximation.
The latter is also true if the cubic terms of the potential vanish.
We put these cases in context and analyse the resulting dynamics,
including a detuning of the $1{:}2{:}4$~resonance within the
particle chain.
\end{abstract}

Key words: Fermi-Pasta-Ulam chain; resonance; normal forms; integrability;

\section{Introduction}
\label{introduction}

\noindent
The Fermi--Pasta--Ulam (\FPU) chain is one of the paradigmatic
examples of a Hamiltonian system, see~\cite{for92, CFB10} and
references therein.
The Hamiltonian function is
\begin{subequations}
\label{HamiltonianFPU}
\begin{align}
   H(p,q) & \;\; = \;\; \sum_{j=1}^N \;
  \left( \frac{p_j^2}{2 m} \, + \, V(q_{j+1} - q_j) \right)
\label{FPUHamiltonian}\\
\intertext{with $N$ the number of particles, mass~$m$ a positive
constant and nearest-neighbour potential}
\label{FPUpotential}
   V(z) & \;\; = \;\; \frac{1}{2}z^2 \; + \;
   \frac{\alpha}{3} z^3 \; + \; \frac{\beta}{4} z^4
   \enspace .
\end{align}
\end{subequations}
The potential~$V$ can be extended to higher powers in~$z$.
In the literature separate attention is often paid to the
$\alpha$--chain ($\beta = 0$) and the $\beta$--chain ($\alpha = 0$).
As an extension of harmonic oscillator interaction the
$\beta$--chain is slightly more natural.

The dynamics defined by the Hamiltonian function~\eqref{HamiltonianFPU}
is described by the equations of motion
\begin{displaymath}
   \dot{q}_j \; = \; \frac{\partial H}{\partial p_j} \; = \;
   \frac{p_j}{m}
   \enspace , \quad
   \dot{p}_j \; = \; - \frac{\partial H}{\partial q_j} \; = \;
   V^{\prime}(q_{j+1} - q_j) \, - \, V^{\prime}(q_j - q_{j-1})
   \enspace , \quad
   j = 1, \ldots, N
\end{displaymath}
for $N$~particles under the force $F = -V^{\prime}$.
The convention $q_{N+1} := q_1$ puts the $N$--degrees-of-freedom (dof)
chain into a circular configuration, one also speaks of the spatially
periodic \FPU-chain.
Using the diagonal $\pS$--symmetry
\begin{equation}
\label{S1symmetry}
\begin{array}{ccc}
   \pS \times \R^{2N} & \longrightarrow & \R^{2N} \\
   (\theta, p, q) & \mapsto & (p ,(q_j + \theta)_j)
\end{array}
\end{equation}
enables us to reduce the equations of motion, thereby fixing the
value of the momentum mapping
\begin{displaymath}
   (p, q) \;\; \mapsto \;\; \sum_{j=1}^N p_j
\end{displaymath}
corresponding with the linear momentum integral.
This leads to a Hamiltonian system with $N-1$~dof.

The initial interest in~\cite{FPU55} was with large~$N$, at that
time $N = 32$ or $N = 64$, trying to confirm the ergodic hypothesis,
see also~\cite{fer23}.
After this failed --- the numerical experiment~\cite{FPU55} did
not show equipartition of energy at low energy level but quite to
the contrary recurrence phenomena --- many papers appeared on the
\FPU-chain and also smaller values of~$N$ were considered.
Interestingly, for $N = 3$ the reduced system in $N - 1 = 2$ dof
was shown in~\cite{for92} to be equivalent to the paradigmatic
H\'enon--Heiles system~\cite{HH64}, a $\Z_3$--symmetric unfolding
of the $1{:}1$~resonance.
Moreover, as shown in~\cite{RV00, rin01} the $1{:}1$~resonance also
governs the normal form approximations of the \FPU-chains with
$N \geq 4$ and their normal forms are all integrable.
This puts the classical \FPU-chain in the realm of
Kolmogorov--Arnol'd--Moser (\kam) theory where invariant maximal
tori of the approximating normal form persist for the original system.
The invariant tori obstruct or at least delay equipartition of energy,
see also~\cite{Zas}.
It turns out that such integrable approximations do also exist for the
\FPU-chain with alternating masses~\cite{GGMV92, BV19}, see in
particular~\cite{BV18} where the case $m_j = 1, m, 1, m$ of four
alternating masses is analysed.

\subsection{Problem formulation}
\label{problemformulation}

In the inhomogenous \FPU-chain the masses of the oscillating
particles introduced in the Hamiltonian~\eqref{FPUHamiltonian}
are not required to be equal.
We follow~\cite{BV17} and consider four masses $m_1, \ldots, m_4$ for
which the fourth and first mass are also subject to nearest-neighbour
interaction, resulting in a spatially periodic \FPU-chain.
Having unequal masses yields $4$~extra parameters --- next to the
coefficients $\alpha, \beta$ of the potential for the
nearest-neighbour interaction --- which can be brought down to one
extra parameter by restricting to mass ratios that produce the
$1{:}2{:}4$~resonance.
Indeed, from~\cite{BV17} we know that there are $12$~curves of mass
ratios producing the $1{:}2{:}4$~resonance.

The analysis of the equations of motion is carried out by computing
normal forms.
The approximations obtained in this way are valid in a neighbourhood
of the equilibrium.
In our problem the origin of the phase space is a stable equilibrium
and, as we shall see, typical error estimates depend on the
size~$\varepsilon$ of a neighbourhood of this equilibrium.

From the approximating normal forms of the Hamiltonian function we
want to identify invariant manifolds and ideally integrability of the
normal form system.
In addition we look for periodic solutions on a given energy
manifold, in some cases normal modes.
We have to explain the terminology as there exists some confusion
in the literature.
In physics a normal mode was originally a continuation of linearized
motion restricted to one of the co-ordinate planes.
However, reductions and transformations of the positions and momenta
mix the variables and the original simple notion of normal mode gets
lost.
In our terminology we will for instance for $n$~actions
$\tau = \tau_1, \ldots, \tau_n$ and corresponding $n$~angles~$\phi$
call $\tau_1$ a normal mode if we have identified a continuation of
a periodic solution of the normal form for which
$\tau_2 = \tau_3 = \ldots = \tau_n = 0$.

The $1{:}2{:}4$~resonance with $\alpha \neq 0$ has in general not an
integrable normal form, see~\cite{chr12}.
However, if two opposite masses in the periodic chain are equal,
then exchanging these masses yields a $\Z_2$--symmetry that enforces
the cubic terms in the normal form to vanish, whereas the quartic
terms produce the first non-trivial normal form.
The same happens for the $\beta$--chain where the third order terms
(with coefficient~$\alpha$) are already zero in the original system.
Such fourth order normal form approximations are integrable.
Interestingly, it turns out that while the quintic terms in the
normal form vanish as well, higher order truncations (starting at
order six) do break the integrability of the normal form.
This yields four different timescales in the dynamics of these
inhomogenous \FPU-chains: the fast harmonic motions in
$1{:}2{:}4$~resonance, the semi-slow motion of the first
non-trivial normal form that can be reduced to $1$~dof, the slow
motion of all higher order normal forms, which still can be
reduced to $2$~dof and the exponentially slow motion in $3$~dof.

As mentioned before, the normal form approximations have validity
near equilibrium, for the approximation theorems see~\cite{SVM07}.
Suppose we consider solutions on an energy manifold near a
{\em stable} equilibrium as is the case in $N-1$~dof for the
Hamiltonian reduced from~\eqref{HamiltonianFPU}.
When considering solutions near a stable equilibrium it is natural
to introduce a small positive parameter~$\varepsilon$ and rescale
$q = \varepsilon \bar{q}$, $p = \varepsilon \bar{p}$, divide the
Hamiltonian by $\varepsilon^2$ and leave out the bars.
In the equations of motion this produces
$\varepsilon \alpha, \varepsilon^2 \beta$ instead of $\alpha, \beta$.
The normalisation procedure keeps the quadratic part $H_2$ of the
Hamiltonian~\eqref{HamiltonianFPU} as an integral of the normal form
equations.

A general result for time-independent Hamiltonians is that when
normalising and using the cubic terms of the Hamiltonian, the
solutions and integrals produce approximations with error
$O(\varepsilon)$ valid on the timescale~$1 / \varepsilon$.
In our special problem where the cubic terms vanish after the
normalising transformation, the error estimate, when including
the normalised quartic terms, is $O(\varepsilon^2)$ valid on the
timescale~$1 / \varepsilon^2$.

The same estimates hold for integral manifolds of normalised systems,
but with a difference: they represent structures that are asymptotic
approximations of phantom structures in the original system.
For the normal form integral~$H_2$ the timescale of validity can be
extended, the approximation is valid for all time as the solutions
in a neighbourhood of equilibrium are bounded.
When including detuning effects, the approximation approach sketched
here enables us to determine the size of the detuning parameters.
When retaining nontrivial cubic terms they have to be of
size~$O(\varepsilon)$; if the normalised cubic terms vanish, the
detuning is bounded by~$O(\varepsilon^2)$.

\subsection{Plan of the paper}
\label{planpaper}

In the next section we formulate the inhomogenous \FPU-chain in more
detail, determine the masses that produce the $1{:}2{:}4$~resonance
and reduce the $\pS$--symmetry~\eqref{S1symmetry} of rotating the
circular chain to arrive at our system in $3$~dof;
classically this means that we use an integral independent of the energy
to remove $1$~dof of the $4$--particles system.
Section~\ref{harmonicoscillators124resonance} then concerns the technical
analysis of determining basic invariants to obtain the normal forms of
the $1{:}2{:}4$~resonance.
In section~\ref{equalmassesFPUchain} we identify the cases with
two equal masses, next to the $\beta$--chain the focus of this paper.
A useful reduction to a $1$~dof problem is analysed in
section~\ref{dynamics1dof}.
Section~\ref{reconstructiondynamics} is about the non-integrable
higher order terms for $2$~dof and the implications for $3$~dof;
we conclude with section~\ref{conclusions}.

\section{The inhomogenous \FPU-chain}
\label{inhomogenousFPUchain}

For the inhomogeneous \FPU-chain we allow the $4$~particle
masses~$m_j$ to be different.
In the Hamiltonian~\eqref{FPUHamiltonian} we take $N=4$ and obtain
\begin{equation}
\label{FPUHam}
   H(p,q) \;\; = \;\; \sum_{j=1}^4 \;
   \left( \frac{p_j^2}{2 m_j} \, + \, V(q_{j+1} - q_j) \right)
\end{equation}
while we keep the potential~$V$ defined in~\eqref{FPUpotential}.
We can write the quadratic part of $H(p, q)$ as
\begin{equation}
\label{FPUQuadHam}
   H_2 \;\; = \;\; \frac{1}{2} p^T A_4 p \; + \; \frac{1}{2} q^T C_4 q
\end{equation}
with $A_4 = \diag(\mu_1, \ldots, \mu_4)$ the $4 \times 4$ diagonal
matrix that has at position $(j, j)$ the inverse masses
$m_j^{-1} =: \mu_j$ while the $4 \times 4$ matrix
\begin{displaymath}
   C_4 \;\; = \;\; \left(
   \begin{array}{cccc}
      2 & -1 & 0 & -1 \\
      -1 & 2 & -1 & 0 \\
      0 & -1 & 2 & -1  \\
      -1 & 0 & -1 & 2
   \end{array}
   \right)
\end{displaymath}
encodes the quadratic part of the potential~$V$.
As shown in~\cite{BV17} one of the eigenvalues of $A_4 C_4$ vanishes,
corresponding to the $\pS$--symmetry~\eqref{S1symmetry} and leading
to the linear momentum integral, while the three remaining eigenvalues
$\gamma_1, \gamma_2, \gamma_3$ of $A_4 C_4$ put the restrictions
\begin{subequations}
\label{MassesFPU}
\begin{eqnarray}
   4(\mu_1 + \mu_3) \mu_2 \mu_4 \; + \; 4(\mu_2 + \mu_4) \mu_1 \mu_3
   & = & \gamma_1 \gamma_2 \gamma_3 \;
\label{MassesFPU1}\\
   3(\mu_1 + \mu_3)(\mu_2 + \mu_4) \; + \; 4(\mu_1\mu_3 + \mu_2\mu_4) & = &
   \gamma_1 \gamma_2 \; + \; \gamma_2 \gamma_3 \; + \; \gamma_1 \gamma_3
\label{MassesFPU2}\\
   2(\mu_1 + \mu_2 + \mu_3 + \mu_4) & = &
   \gamma_1 \; + \; \gamma_2 \; + \; \gamma_3
\label{MassesFPU3}
\end{eqnarray}
\end{subequations}
on the inverse masses~$\mu_j$.
Correspondingly, choices of the~$\mu_j$ exist that produce the first
order resonances $1{:}2{:}1$, $1{:}2{:}3$ and~$1{:}2{:}4$ (but
{\em not} $1{:}2{:}2$).
The resulting $1{:}2{:}3$~resonance has been studied in~\cite{BV17},
here we consider the $1{:}2{:}4$~resonance.
Scaling to $\gamma_1 + \gamma_2 + \gamma_3 + \gamma_4 = 21$ we
have $\gamma_1 = 1$, $\gamma_2 = 4$ and $\gamma_3 = 16$ next to
$\gamma_4 = 0$ whence the right hand side $\gamma_1 \gamma_2 \gamma_3$
of eq.~\eqref{MassesFPU1} becomes~$64$ and the right hand side of
eq.~\eqref{MassesFPU2} becomes~$84$.
We use $\mu_2 + \mu_4 =: v$ to parametrise the solutions
$(\mu_1, \mu_2, \mu_3, \mu_4) \in \R^4_{>0}$ and first note that
$v \neq \frac{21}{4}$.
Indeed otherwise also $\mu_1 + \mu_3 = \frac{21}{4}$ by
eq.~\eqref{MassesFPU3}, whence
$\mu_1 \mu_3 + \mu_2 \mu_4 = \frac{21}{64}$ by eq.~\eqref{MassesFPU2}
and $\mu_1 \mu_3 + \mu_2 \mu_4 = \frac{64}{21}$ by
eq.~\eqref{MassesFPU1} contradict each other.
Then the solution of these equations is
\begin{eqnarray*}
   \mu_{1,3} & = & \frac{21 - 2 v \pm \sqrt{\Delta}}{4} \\
   \mu_{2,4} & = & \frac{v \pm \sqrt{\Gamma}}{2}
\end{eqnarray*}
where
\begin{eqnarray*}
   \Delta & = & - \frac{(2 v + 11)(2 v - 19)(2 v - 13)}{4 v - 21} \\
   \Gamma & = & - \frac{2 (v - 16)(v - 4)(v - 1)}{4 v - 21}
   \enspace .
\end{eqnarray*}
Diagonalising
$A_4^{\frac{1}{2}} C_4 A_4^{\frac{1}{2}} = U \Lambda U^T$,
$\Lambda = \diag(\gamma_1, \gamma_2, \gamma_3, 0)$ with an
orthogonal matrix~$U$ yields the symplectic transformation
\begin{equation}
\label{transformationFPU}
   p \; = \; K y \enspace , \quad q \; = \; L x
\end{equation}
adapted from~\cite{BV17} where
\begin{displaymath}
   K \; = \; A_4^{-\frac{1}{2}} U \Omega^{\frac{1}{4}}
   \quad \mbox{and} \quad
   L \; = \; A_4^{\frac{1}{2}} U \Omega^{-\frac{1}{4}}
\end{displaymath}
with $\Omega = \diag(\gamma_1, \gamma_2, \gamma_3, 1)$ that turns the
quadratic part~\eqref{FPUQuadHam} of the Hamiltonian~\eqref{FPUHam}
into
\begin{equation}
\label{FPUQuad}
   H_2 \;\; = \;\; \sum_{j=1}^3 2^{j-1} (\frac{x_j^2 + y_j^2}{2})
   \; + \; \frac{y_4^2}{2} \enspace .
\end{equation}
This simultaneously achieves two goals.
On one hand the $\pS$--action of eq.~\eqref{S1symmetry} becomes
$x_4 \mapsto x_4 + \vartheta$, i.e.\ $x_4$ is a cyclic angle and
reducing this $\pS$--symmetry amounts to fixing the value~$y_4$
of the momentum integral and ignoring the cyclic angle~$x_4$,
whence the reduced Hamiltonian in $3$~dof reads as
\begin{equation}
\label{Quad}
   H_2 \;\; = \;\; \frac{x_1^2 + y_1^2}{2} \; + \;
   2 (\frac{x_2^2 + y_2^2}{2}) \; + \; 4 (\frac{x_3^2 + y_3^2}{2})
   \enspace .
\end{equation}
On the other hand the transformation~\eqref{transformationFPU}
achieves the splitting of the quadratic part of the reduced
Hamiltonian into the three oscillators in $1{:}2{:}4$~resonance
visible in eq.~\eqref{Quad}.
For inverse masses~$\mu_j$ that do not lead to resonant oscillators,
this same procedure --- with adapted $\Lambda$ and~$\Omega$ ---
leads to the deformation
\begin{equation}
\label{Quadlam}
   H_2^{\lambda} \;\; = \;\; (1 + \lambda_1) \frac{x_1^2 + y_1^2}{2}
   \; + \; (2 + \lambda_2) \frac{x_2^2 + y_2^2}{2} \; + \;
   (4 + \lambda_3) \frac{x_3^2 + y_3^2}{2}
\end{equation}
of the $1{:}2{:}4$~resonant oscillator~\eqref{Quad}, with detunings
$\lambda_j = \lambda_j(\mu)$, $j = 1, 2, 3$.
This is the quadratic part of
\begin{equation}
\label{Hamlam}
   H^{\lambda}(x, y) \;\; = \;\; H(Ky, Lx)
\end{equation}
reduced from~\eqref{FPUHam}.
As we have discussed in section~\ref{problemformulation},
the detunings $\lambda_1, \lambda_2, \lambda_3$ depend in size on the
order of the first non-zero terms in the normal form.

\section{Harmonic and nonlinear oscillators in $1{:}2{:}4$ resonance}
\label{harmonicoscillators124resonance}

It is well-known that the flow induced by the quadratic
Hamiltonian~\eqref{Quad} is periodic and a truncated normal form of
the Hamiltonian~\eqref{Hamlam} with respect to~\eqref{Quad} depends
on $x, y$ only through certain invariants, see below (for the general
theory see also chs.\ 11--13 in~\cite{SVM07}).
This allows to use the periodic flow of~$\X{H_2^0}$ to reduce to
$2$~dof.

This embeds the reduced system into the space of invariants~$\R^{11}$
where $11$ is the number of independent invariants, see below.
However, as the $11 \times 11$ Poisson matrix of these invariants
has rank~$4$, this is indeed a reduction from $3$ to $2$~dof
(the Hamiltonian~\eqref{Quad} acts as a Casimir).
The effectiveness of determining invariants becomes clear in
section~\ref{equationsmotion} where we formulate the normalised
equations of motion.

Note that a preliminary study of the Hamiltonian~\eqref{Hamlam} at
$\lambda = 0$ is contained in~\cite{EvdA83}, for a summary
see~\cite{SVM07}.
We will pursue the study of such non-integrable normal forms in a
subsequent paper.
As shown in~\cite{chr12} already the general third order normal
form of the $1{:}2{:}4$ resonance is not integrable.

In case there does exist a third independent integral (next to the
quadratic~\eqref{Quad} and the Hamiltonian itself) the normalised
system is integrable and we can reduce to $1$~dof.
As noted in~\cite{SVM07} this may happen if there are additional
discrete symmetries.
Such a discrete symmetry is introduced if two opposite masses are
equal as one then can flip the periodic chain about the other two
masses without changing the equations of motion.

\subsection{Invariants, syzygies and Poisson bracket relations}
\label{invariantssyzygiesPoissonbracketrelations}

In complex co-ordinates $z_k = x_k + \I y_k$ the flow defined
by~$H_2^0$ reads as
\begin{displaymath}
   (t, z) \;\; \mapsto \;\; \left(
   \E^{- \I t} z_1, \E^{- 2 \I t} z_2, \E^{- 4 \I t} z_3 \right)
\end{displaymath}
whence a monomial
$z_1^{m_1} z_2^{m_2} z_3^{m_3} \bar{z}_1^{n_1} \bar{z}_2^{n_2}
 \bar{z}_3^{n_3}$
is invariant if and only if
$m_1 + 2 m_2 + 4 m_3 = n_1 + 2 n_2 + 4 n_3$.
The ring of functions invariant under this flow is generated by
the basic invariants
\begin{eqnarray*}
   \tau_1 & = & \frac{1}{2} z_1 \bar{z}_1
   \;\; = \;\; \frac{x_1^2 + y_1^2}{2} \\
   \tau_2 & = & \frac{1}{2} z_2 \bar{z}_2
   \;\; = \;\; \frac{x_2^2 + y_2^2}{2} \\
   \tau_3 & = & \frac{1}{2} z_3 \bar{z}_3
   \;\; = \;\; \frac{x_3^2 + y_3^2}{2}
\end{eqnarray*}
of degree~$2$, the basic invariants
\begin{eqnarray*}
   \sigma_1 & = & \frac{1}{2} \re z_1^2 \bar{z}_2
   \;\; = \;\; \frac{x_1^2 x_2 + 2 x_1 y_1 y_2 - x_2 y_1^2}{2} \\
   \sigma_2 & = & \frac{1}{2} \im z_1^2 \bar{z}_2
   \;\; = \;\; \frac{y_1^2 y_2 + 2 x_1 x_2 y_1 - x_1^2 y_2}{2} \\
   \sigma_3 & = & \frac{1}{2} \re z_2^2 \bar{z}_3
   \;\; = \;\; \frac{x_2^2 x_3 + 2 x_2 y_2 y_3 - x_3 y_2^2}{2} \\
   \sigma_4 & = & \frac{1}{2} \im z_2^2 \bar{z}_3
   \;\; = \;\; \frac{y_2^2 y_3 + 2 x_2 x_3 y_2 - x_2^2 y_3}{2}
\end{eqnarray*}
of degree~$3$, the basic invariants
\begin{eqnarray*}
   \sigma_5 & = & \frac{1}{2} \re z_1^2 z_2 \bar{z}_3
   \;\; = \;\; \frac{x_1^2 (x_2 x_3 + y_2 y_3) \, + \, 2 x_1 y_1
   (x_2 y_3 - x_3 y_2) \, - \, y_1^2(x_2 x_3 + y_2 y_3)}{2} \\
   \sigma_6 & = & \frac{1}{2} \im z_1^2 z_2 \bar{z}_3
   \;\; = \;\; \frac{-x_1^2 (x_2 y_3 - x_3 y_2) \, + \, 2 x_1 y_1
   (x_2 x_3 + y_2 y_3) \, + \, y_1^2 (x_2 y_3 - x_3 y_2)}{2}
\end{eqnarray*}
of degree~$4$ and the basic invariants
\begin{eqnarray*}
   \sigma_7 & = & \frac{1}{24} \re z_1^4 \bar{z}_3
   \;\; = \;\; \frac{(x_1^4 - 6 x_1^2 y_1^2 + y_1^4) x_3 \, + \,
   4 (x_1^2 - y_1^2) x_1 y_1 y_3}{24} \\
   \sigma_8 & = & \frac{1}{24} \im z_1^4 \bar{z}_3
   \;\; = \;\; \frac{4 (x_1^2 - y_1^2) x_1 x_3 y_1 \, - \,
   (x_1^4 - 6 x_1^2 y_1^2 + y_1^4) y_3}{24}
\end{eqnarray*}
of degree~$5$.
\begin{table}
\begin{center}
\captionof{table}{
   Poisson brackets $\{ \tau_i, \sigma_j \}$
   \label{table1}
}
\begin{tabular}{c|cccccccc}
   $\{ \downarrow , \rightarrow \}$ &
   $\sigma_1$ & $\sigma_2$ & $\sigma_3$ & $\sigma_4$ & $\sigma_5$ &
   $\sigma_6$ & $\sigma_7$ & $\sigma_8$ \\ \hline
   $\tau_1$ & $-2 \sigma_2$ & $2 \sigma_1$ & $0$ &
   $0$ & $-2 \sigma_6$ & $2 \sigma_5$ & $-4 \sigma_8$ & $4 \sigma_7$ \\
   $\tau_2$ & $\sigma_2$ & $-\sigma_1$ &
   $-2 \sigma_4$ & $2 \sigma_3$ & $-\sigma_6$ & $\sigma_5$ & $0$ & $0$ \\
   $\tau_3$ & $0$ & $0$ & $\sigma_4$ & $-\sigma_3$ &
   $\sigma_6$ & $-\sigma_5$ & $\sigma_8$ & $-\sigma_7$
\end{tabular}
\end{center}
\end{table}
These $11$~basic invariants\footnote{These are the generators given
in~\cite{HH07}, note however that only $9$ of them are of
degree~$\leq 4$.} are not free, but restricted by the syzygies
$S_j = 0$, $j = 1, \ldots, 20$ given by
\begin{eqnarray*}
   S_1 & = & \frac{1}{2} (\sigma_1^2 + \sigma_2^2) \; - \;
   \tau_1^2 \tau_2 \\
   S_2 & = & \frac{1}{2} (\sigma_3^2 + \sigma_4^2) \; - \;
   \tau_2^2 \tau_3 \\
   S_3 & = & \sigma_1 \sigma_3 - \sigma_2 \sigma_4 \; - \;
   \tau_2 \sigma_5 \\
   S_4 & = & \sigma_2 \sigma_3 + \sigma_1 \sigma_4 \; - \;
   \tau_2 \sigma_6 \\
   S_5 & = & \sigma_1 \sigma_5 + \sigma_2 \sigma_6 \; - \;
   2 \tau_1^2 \sigma_3 \\
   S_6 & = & \sigma_1 \sigma_6 - \sigma_2 \sigma_5 \; - \;
   2 \tau_1^2 \sigma_4 \\
   S_7 & = & \sigma_1 \sigma_5 - \sigma_2 \sigma_6 \; - \;
   12 \tau_2 \sigma_7 \\
   S_8 & = & \sigma_2 \sigma_5 + \sigma_1 \sigma_6 \; - \;
   12 \tau_2 \sigma_8 \\
   S_9 & = & \sigma_3 \sigma_5 + \sigma_4 \sigma_6 \; - \;
   2 \tau_2 \tau_3 \sigma_1 \\
   S_{10} & = & \sigma_3 \sigma_6 - \sigma_4 \sigma_5 \; - \;
   2 \tau_2 \tau_3 \sigma_2 \\
   S_{11} & = & \frac{1}{4} (\sigma_5^2 + \sigma_6^2) \; - \;
   \tau_1^2 \tau_2 \tau_3 \\
   S_{12} & = & 6 (\sigma_1 \sigma_7 + \sigma_2 \sigma_8) \; - \;
   \tau_1^2 \sigma_5 \\
   S_{13} & = & 6 (\sigma_1 \sigma_8 - \sigma_2 \sigma_7) \; - \;
   \tau_1^2 \sigma_6 \\
   S_{14} & = & 6 (\sigma_3 \sigma_7 - \sigma_4 \sigma_8) \; - \;
   \frac{1}{2} (\sigma_5^2 - \sigma_6^2) \\
   S_{15} & = & 6 (\sigma_3 \sigma_8 + \sigma_4 \sigma_7) \; - \;
   \sigma_5 \sigma_6 \\
   S_{16} & = & 3 (\sigma_3 \sigma_7 + \sigma_4 \sigma_8) \; - \;
   \frac{1}{2} \tau_3 (\sigma_1^2 - \sigma_2^2) \\
   S_{17} & = & 3 (\sigma_3 \sigma_8 - \sigma_4 \sigma_7) \; - \;
  \tau_3 \sigma_1 \sigma_2\\
   S_{18} & = & 3 (\sigma_5 \sigma_7 + \sigma_6 \sigma_8) \; - \;
   \tau_1^2 \tau_3 \sigma_1 \\
   S_{19} & = & 3 (\sigma_5 \sigma_8 - \sigma_6 \sigma_7) \; - \;
  \tau_1^2 \tau_3 \sigma_2 \\
   S_{20} & = & 18 (\sigma_7^2 + \sigma_8^2) \; - \; \tau_1^4 \tau_3
\end{eqnarray*}
and by the inequalities $\tau_1 \geq 0$, $\tau_2 \geq 0$,
$\tau_3 \geq 0$.
Note that the syzygies themselves have to satisfy (at least)
$14$~relations.
The Poisson bracket relations between the basic invariants are
\begin{displaymath}
   \{ \tau_i, \tau_j \} \;\; = \;\; 0
   \enspace , \quad
   i, j \; = \; 1, 2, 3
\end{displaymath}
while the $\{ \tau_i, \sigma_j \}$, $i = 1, 2, 3$, $j = 1, \ldots, 8$
are given in table~\ref{table1}.
The relations $\{ \sigma_i, \sigma_j \}$, $i, j = 1, \ldots, 8$
are split into tables \ref{table2} and~\ref{table3} (here we use
that $\{ \sigma_i, \sigma_j \} = - \{ \sigma_j, \sigma_i \}$).

\begin{table}
\begin{center}
\captionof{table}{
   Poisson brackets $\{ \sigma_i, \sigma_j \}$
   \label{table2}
}
\begin{tabular}{c|cccc}
   $\{ \downarrow , \rightarrow \}$ &
   $\sigma_1$ & $\sigma_2$ & $\sigma_3$ & $\sigma_4$  \\ \hline
   $\sigma_1$ & $0$ & $-\tau_1(\tau_1-4\tau_2)$ & $-\sigma_6$ & $\sigma_5$  \\
   $\sigma_2$ & $\tau_1(\tau_1-4\tau_2)$ & $0$ & $\sigma_5$ & $\sigma_6$  \\
   $\sigma_3$ & $\sigma_6$ & $-\sigma_5$ & $0$ & $-\tau_2(\tau_2-4\tau_3)$  \\
   $\sigma_4$ & $-\sigma_5$ & $-\sigma_6$ & $\tau_2(\tau_2-4\tau_3)$ & $0$  \\
   $\sigma_5$ & $4\tau_1\sigma_4+6\sigma_8$ & $4\tau_1\sigma_3-6\sigma_7$ & $-\sigma_2(\tau_2-2\tau_3)$ & $-\sigma_1(\tau_2-2\tau_3)$  \\
   $\sigma_6$ & $-4\tau_1\sigma_3-6\sigma_7$ & $4\tau_1\sigma_4-6\sigma_8$ & $\sigma_1(\tau_2-2\tau_3)$ & $-\sigma_2(\tau_2-2\tau_3)$  \\
   $\sigma_7$ & $\frac{2}{3}\tau_1\sigma_6$ & $\frac{2}{3}\tau_1\sigma_5$ & $-\frac{1}{6}\sigma_1\sigma_2$ & $-\frac{1}{12}(\sigma_1^2-\sigma_2^2)$  \\
   $\sigma_8$ & $-\frac{2}{3}\tau_1\sigma_5$ & $\frac{2}{3}\tau_1\sigma_6$ & $\frac{1}{12}(\sigma_1^2-\sigma_2^2)$ & $-\frac{1}{6}\sigma_1\sigma_2$
\end{tabular}
\end{center}
\end{table}

The Poisson bracket relations together with the syzygies allow to
identify the subsets $\{ \tau_1 = 0 \}$ and $\{ \tau_1 > 0 \}$ of
the reduced phase space that are invariant under the dynamics for
every Hamiltonian function~$\cH$ that can be written in terms of
the invariants, in particular for truncated normal forms.
Indeed, for $\tau_1 = 0$ the syzygies $S_1$, $S_{11}$ and~$S_{20}$
enforce
$\sigma_1 = \sigma_2 = \sigma_5 = \sigma_6 = \sigma_7 = \sigma_8 = 0$
whence
\begin{displaymath}
   \dot{\tau}_1 \;\; = \;\;
    - 2 \sigma_2 \frac{\partial \cH}{\partial \sigma_1} \; + \;
    2 \sigma_1 \frac{\partial \cH}{\partial \sigma_2} \; - \;
    2 \sigma_6 \frac{\partial \cH}{\partial \sigma_5} \; + \;
    2 \sigma_5 \frac{\partial \cH}{\partial \sigma_6} \; - \;
    4 \sigma_8 \frac{\partial \cH}{\partial \sigma_7} \; + \;
    4 \sigma_7 \frac{\partial \cH}{\partial \sigma_8} \;\; = \;\; 0
\end{displaymath}
and similarly
$\dot{\sigma}_1 = \dot{\sigma}_2 = \dot{\sigma}_5 = \dot{\sigma}_6
 = \dot{\sigma}_7 = \dot{\sigma}_8 = 0$
--- independent of~$\cH$ --- ensure that the values of these
invariants, in particular of~$\tau_1$, stay zero.
This makes the complement $\{ \tau_1 > 0 \}$ an invariant set as
well.

This does not work for $\tau_2 = 0$ because of the
$\tau_1^2$--contribution from $\{ \sigma_1, \sigma_2 \}$
which furthermore prevents equilibria\footnote{The $3$~normal modes
are periodic orbits in $3$~dof and reduce to equilibria in $2$~dof.
In the literature one sometimes speaks of `relative equilibria' to
allude to this fact, but here we have made the symmetry reduction
from $3$~dof to $2$~dof explicit and simply speak of `equilibria'
when arguing in $2$~dof (i.e.\ in the $\tau, \sigma$ variables).}
inside $\{ \tau_2 = 0, \tau_1 > 0 \}$.
Even if $\cH$ is independent of $\sigma_1$ and~$\sigma_2$ there are
non-zero terms in e.g.\ the brackets $\{ \sigma_1, \sigma_5 \}$,
$\{ \sigma_5, \sigma_6 \}$ and $\{ \sigma_7, \sigma_8 \}$.
Similarly for $\tau_3 = 0$ because of the $\tau_2^2$--contribution
from $\{ \sigma_3, \sigma_4 \}$, also preventing equilibria inside
$\{ \tau_3 = 0, \tau_1 > 0 \}$.

Furthermore this shows that the normal--$1$--mode
$(\tau, \sigma) = (\tau_1, 0, \ldots, 0)$
is not automatically an equilibrium;
similarly for the normal--$2$--mode
$(\tau, \sigma) = (0, \tau_2, 0, \ldots, 0)$.
However, the normal--$3$--mode
$(\tau, \sigma) = (0, 0, \tau_3, 0, \ldots, 0)$
makes all Poisson brackets vanish, whence it is an equilibrium
for every Hamiltonian function.
This can also be obtained from the dynamics on the invariant set
$\{ \tau_1 = 0 \}$, which is best understood by intersecting the
energy level set with the surface of revolution
\begin{displaymath}
   \sigma_3^2 \; + \; \sigma_4^2
   \;\; = \;\; \tau_2^2 \, (\eta - \tau_2)
   \enspace , \quad
   0 \leq \tau_2 \leq \eta
\end{displaymath}
defined by the syzygy $S_2 = 0$.
Here $\eta \geq 0$ is the value of the conserved
quantity~$\frac{1}{2} H_2^0$, allowing to replace $\tau_3$ by
$\frac{1}{2}(\eta - \tau_2)$ since $\tau_1 = 0$.
The two surfaces touch in (regular) equilibria, while the singular
point $\tau_2 = 0$ --- the normal $3$--mode --- is always an
equilibrium.

\begin{table}
\begin{center}
\captionof{table}{
   Poisson brackets $\{ \sigma_i, \sigma_j \}$ (continued)
   \label{table3}
}
\begin{tabular}{c|cccc}
   $\{ \downarrow , \rightarrow \}$ &
   $\sigma_5$ & $\sigma_6$ & $\sigma_7$ & $\sigma_8$ \\ \hline
   $\sigma_5$ & $0$ & $\!\!\!\!\!\!\!\!-2\tau_1(\tau_1\tau_2-\tau_1\tau_3-4\tau_2\tau_3)$ & $\frac{1}{6}\tau_1\sigma_2(\tau_1-8\tau_3)$ & $-\frac{1}{6}\tau_1\sigma_1(\tau_1-8\tau_3)$ \\
   $\sigma_6$ & $2\tau_1(\tau_1\tau_2-\tau_1\tau_3-4\tau_2\tau_3)\!\!\!\!\!\!\!\!$ & $0$ & $\frac{1}{6}\tau_1\sigma_1(\tau_1-8\tau_3)$ & $\frac{1}{6}\tau_1\sigma_2(\tau_1-8\tau_3)$ \\
   $\sigma_7$ & $-\frac{1}{6}\tau_1\sigma_2(\tau_1-8\tau_3)$ & $-\frac{1}{6}\tau_1\sigma_1(\tau_1-8\tau_3)$ & $0$ & $-\frac{1}{36}\tau_1^3(\tau_1-16\tau_3)$\\
   $\sigma_8$ & $\frac{1}{6}\tau_1\sigma_1(\tau_1-8\tau_3)$ & $-\frac{1}{6}\tau_1\sigma_2(\tau_1-8\tau_3)$ & $\frac{1}{36}\tau_1^3(\tau_1-16\tau_3)$ & $0$
\end{tabular}
\end{center}
\end{table}

On the invariant subset $\{ \tau_1 > 0 \}$ of the reduced phase space
in $2$~dof we may use the syzygies to replace
\begin{eqnarray*}
   \tau_2 & = & \frac{\sigma_1^2 + \sigma_2^2}{2 \tau_1^2} \\
   \tau_3 & = & 18 \, \frac{\sigma_7^2 + \sigma_8^2}{\tau_1^4} \\
   \sigma_3 & = & \frac{6 \sigma_1 \sigma_2 \sigma_8 \, + \,
   3 (\sigma_1^2 - \sigma_2^2) \sigma_7}{\tau_1^4} \\
   \sigma_4 & = &
   \frac{3 (\sigma_1^2 - \sigma_2^2) \sigma_8 \, - \,
   6 \sigma_1 \sigma_2 \sigma_7}{\tau_1^4} \\
   \sigma_5 & = &
   6 \, \frac{\sigma_1 \sigma_7 + \sigma_2 \sigma_8}{\tau_1^2} \\
   \sigma_6 & = &
   6 \, \frac{\sigma_1 \sigma_8 - \sigma_2 \sigma_7}{\tau_1^2}
   \enspace ,
\end{eqnarray*}
compare with~\cite{SW16}.
Substituting these expressions in the equations of the remaining
$5$~invariants yields a system of $5$~differential equations where
furthermore the Casimir~$H_2^0$ makes sure that we are in fact on
a $4$--dimensional manifold.
This confirms that we have indeed reduced to $2$~dof.
The reduced system is straightforwardly studied on the singular part
$\{ \tau_1 = 0 \}$ of the reduced phase space, a $2$--dimensional
surface of revolution that is conically attached to the regular part
$\{ \tau_1 > 0 \}$ --- the latter has a simple structure (it is a
manifold) but has the more complicated dynamics of truly $2$~dof.

\subsection{Equations of motion}
\label{equationsmotion}

Any Hamiltonian function on~$\R^6$ that is invariant under the
$\pS$--action generated by the quadratic Hamiltonian~\eqref{Quad}
--- e.g.\ a truncated normal form of the general
Hamiltonian~\eqref{Hamlam} with respect to~\eqref{Quad} --- can be
expressed as a function
\begin{displaymath}
   \cH \;\; = \;\; \cH(\tau, \sigma)
\end{displaymath}
in the basic invariants.
The equations of motion on the reduced phase space then read as
\begin{align*}
   \dot{\tau}_i & \;\; = \;\; \{ \tau_i, \cH \}
   \enspace , \quad i = 1, 2, 3
\vspace*{-19pt}
\intertext{and}
\vspace*{-19pt}
   \dot{\sigma}_j & \;\; = \;\; \{ \sigma_j, \cH \}
   \enspace , \quad j = 1, \ldots, 8.
\end{align*}
We expand $\cH$ as
\begin{equation}
\label{Hamred}
   \cH^{\lambda} \;\; = \;\; \cH_2^{\lambda} \; + \;
   \cH_3 \; + \; \cH_4 \; + \; h.o.t.
\end{equation}
with
\begin{subequations}
\begin{align}
   H_2^{\lambda} & \;\; = \;\; (1 + \lambda_1) \tau_1 \; + \;
   (2 + \lambda_2) \tau_2 \; + \; (4 + \lambda_3) \tau_3
\label{Hamred2}\\
   \cH_3 & \;\; = \;\; a_1 \sigma_1 \; + \; a_2 \sigma_2
   \; + \; a_3 \sigma_3 \; + \; a_4 \sigma_4
\label{Hamred3}
\vspace*{-19pt}
\intertext{and}
\vspace*{-19pt}
   \cH_4 & \;\; = \;\; \frac{1}{2} \sum_{k=1}^3 b_k \tau_k^2
   \; + \; b_4 \sigma_5 \; + \; b_5 \sigma_6
   \; + \; b_6 \tau_2 \tau_3 \; + \; b_7 \tau_1 \tau_3
   \; + \; b_8 \tau_1 \tau_2  \enspace .
\label{Hamred4}
\end{align}
\end{subequations}
Note that the normal form of a Hamiltonian with a positional force
(derived from a potential) as in~\eqref{Hamlam} reduced from the
\FPU-chain~\eqref{FPUHam} has $a_2 = a_4 = 0$ and $b_5 = 0$.
The non-integrability~\cite{chr12} of the general truncated
normal form of order~$3$ is reflected by $(a_1, a_3) \in \R^2$ in
general position.
Discrete symmetries~\cite{SVM07} can lead to integrable normal
forms.

\subsection{Symmetry reduction to one degree of freedom}
\label{symmetryreduction1dof}

Let the Hamiltonian~\eqref{Hamred} be invariant under the
$\Z_2$--symmetry generated by
\begin{displaymath}
   (x_1, x_2, x_3, y_1, y_2, y_3) \;\; \mapsto \;\;
   (x_1, x_2, -x_3, y_1, y_2, -y_3)  \enspace .
\end{displaymath}
This symmetry leaves the $\tau_i$ and $\sigma_1$, $\sigma_2$
invariant while the $\sigma_j$, $j = 3, \ldots, 8$ are mapped
to~$-\sigma_j$.
Consequently, the Hamiltonian~$\cH^{\lambda}$ depends only on even
powers in $\sigma_j$, $j = 3, \ldots, 8$ and in particular
$a_3 = a_4 = 0$ in eq.~\eqref{Hamred3} for~$\cH_3$.
This makes $\tau_3$ an integral of motion of the truncation
$\cH^{3, \lambda} = \cH_2^{\lambda} + \cH_3$ of~$\cH^{\lambda}$.
The flow of~$\X{\tau_3}$ rotates the $(\sigma_3, \sigma_4)$--,
$(\sigma_5, \sigma_6)$-- and $(\sigma_7, \sigma_8)$--planes, leaving
only $\tau_1$, $\tau_2$, $\tau_3$, $\sigma_1$ and~$\sigma_2$
invariant.
Reducing the $\pS$--action generated by~$\tau_3$ leads to $1$~dof
and on the reduced phase space these $5$~invariants can serve as
variables.
The syzygy $S_1 = 0$, the value $\eta$ of~$\frac{1}{2} \cH_2^0$ and
the value $\zeta \leq \half \eta$ of~$\tau_3$ reveal the reduced
phase space to be the surface of revolution
\begin{displaymath}
   \sigma_1^2 \, + \, \sigma_2^2 \; = \;
   8 (\eta - 2 \zeta - \tau_2)^2 \tau_2
   \enspace , \quad 0 \leq \tau_2 \leq \eta - 2 \zeta
   \enspace .
\end{displaymath}
Rotating the $(\sigma_1, \sigma_2)$--plane to achieve $a_2 = 0$ and
omitting constant terms in~$\cH^{3, \lambda}$ we obtain
\begin{displaymath}
   \cH_{\eta, \zeta}^{3, \lambda} \;\; = \;\;
   a_1 \sigma_1 \; + \; (2 \lambda_1 - \lambda_2) \tau_2
   \enspace .
\end{displaymath}
Intersecting the reduced phase space with the planes
$\{ \cH_{\eta, \zeta}^{3, \lambda} = h \}$
then yields the orbits in $1$~dof.
The reconstructed dynamics in $2$~dof persists under
both the perturbation by the quartic part~\eqref{Hamred4} and by
the fifth order part~$\cH_5$.
Indeed, remarkably enough, one has to go to sixth order where
$\cH_6$ breaks the integrability because it does depend on even
powers in $\sigma_3$ and~$\sigma_4$.

\begin{Remark}
This analysis remains true {\it mutatis mutandis} if the
Hamiltonian~\eqref{Hamred} is invariant under the $\Z_2$--symmetry
generated by
\begin{displaymath}
   (x_1, x_2, x_3, y_1, y_2, y_3) \;\; \mapsto \;\;
   (x_1, -x_2, x_3, y_1, -y_2, y_3)  \enspace .
\end{displaymath}
Indeed, this symmetry leaves the $\tau_i$ and $\sigma_3$, $\sigma_4$,
$\sigma_7$, $\sigma_8$ invariant, while $\sigma_1$, $\sigma_2$,
$\sigma_5$, $\sigma_6$ are mapped to $-\sigma_1$, $-\sigma_2$,
$-\sigma_5$ and $-\sigma_6$, respectively.
This makes $\tau_1$ an integral of motion of
$\cH_2^{\lambda} + \cH_3$ which persists under the perturbation
by the quartic part~\eqref{Hamred4}, but in this case already the
fifth order part~$\cH_5$ breaks the integrability.
\end{Remark}

\noindent
In the case that the cubic terms~\eqref{Hamred3} vanish identically,
e.g.\ when the Hamiltonian~\eqref{Hamred} is invariant under the
$\Z_2$--symmetry generated by
\begin{equation}
\label{Z2symmetry}
   (x_1, x_2, x_3, y_1, y_2, y_3) \;\; \mapsto \;\;
   (x_1, -x_2, -x_3, y_1, -y_2, -y_3)  \enspace ,
\end{equation}
we immediately consider the truncation
\begin{displaymath}
   \cH^{4, \lambda} \;\; = \;\; \cH_2^{\lambda} \; + \; \cH_4
\end{displaymath}
given by the fourth order term~\eqref{Hamred4}.
Requesting $b_4^2 + b_5^2 \neq 0$ we can again achieve $b_5 = 0$ by
a rotation, but now the term $b_4 \sigma_5$ in the truncated normal
form is accompanied by a quadratic expression in $\tau_1$, $\tau_2$,
$\tau_3$.
Since $\sigma_1$, $\sigma_2$, $\sigma_3$, $\sigma_4, \sigma_7$
and~$\sigma_8$ do not enter the Hamiltonian $\cH^{4, \lambda}$,
there is a third integral of motion $\cK = \tau_2 + \tau_3$
(or, equivalently, $\tau_1 - 2 \tau_2$, $\tau_1 + 2 \tau_3$ or
$\tau_1 - \tau_2 + \tau_3$).
The flow of~$\X{\cK}$ rotates the $(\sigma_1, \sigma_2)$--,
$(\sigma_3, \sigma_4)$-- and $(\sigma_7, \sigma_8)$--planes, leaving
only $\tau_1$, $\tau_2$, $\tau_3$, $\sigma_5$ and~$\sigma_6$
invariant.
Reducing the $\pS$--action generated by~$\cK$ leads to $1$~dof and on
the reduced phase space these $5$~invariants can serve as variables.
The syzygy $S_{11} = 0$, the value $\eta$ of~$\frac{1}{2} \cH_2^0$
and the value $\zeta \leq \eta$ of~$\cK$ reveal the reduced phase
space to be the surface of revolution
\begin{equation}
\label{redphaseineq}
   \sigma_5^2 \, + \, \sigma_6^2 \; = \;
   4 \tau_1^2 \tau_2 \tau_3 \; = \;
   16 (2 \eta - 4 \zeta + \tau_2)^2 \tau_2 (\zeta - \tau_2)
   \enspace , \quad
   \max(0, 2 \zeta - \eta) \leq \tau_2 \leq \zeta
   \enspace .
\end{equation}
Omitting constant terms, the truncated normal form of order~$4$
becomes
\begin{equation}
\label{redHam}
\begin{array}{rcl}
   \cH_{\eta, \zeta}^{4, \lambda} & = & b_4 \sigma_5 \; + \;
   (4 b_1 + b_2 + b_3 - 2 b_6 + 4 b_7 - 4 b_8)
   \displaystyle{\frac{\tau_2^2}{2}} \\
   & & \!\!\!\! \mbox{} + \; \left[
   2 \lambda_1 + \lambda_2 - \lambda_3 \, + \, 4 b_1 (\eta - 2 \zeta)
   \, - \, b_3 \zeta \right. \\
   & & \mbox{} \left. \quad + \, b_6 \zeta \, + \,
   4 b_7 (\eta - 2 \zeta) \, - \, 2 b_8 (\eta - 3 \zeta) \right]
   \tau_2
   \enspace .
\end{array}
\end{equation}
Intersecting the reduced phase space with the parabolic cylinders
$\{ \cH_{\eta, \zeta}^{4, \lambda} = h \}$
then yields the orbits in $1$~dof.
This is carried out in detail in section~\ref{dynamics1dof} below.
While the fifth order normal form terms~$\cH_5$ vanish, the
perturbation by the sixth order normal form terms~$\cH_6$
generically breaks the integrability and the resulting perturbation
problem has to be analysed in $2$~dof.

In the special case $b_4^2 + b_5^2 = 0$ the dynamics in $1$~dof is
given by periodic orbits around the $\tau_2$--axis, with equilibria
at the extremal values of~$\tau_2$.
In $3$~dof the latter yield invariant $2$--tori
shrinking down to normal modes.
Only where the vertical planes
$\{ \cH_{\eta, \zeta}^{4, \lambda} = h \}$
are double planes intersecting the reduced phase space, i.e.\ where
\begin{displaymath}
   \tau_2 \;\; = \;\; - \frac{2 \lambda_1 + \lambda_2 - \lambda_3
   \, + \, 4 b_1 (\eta - 2 \zeta) \, - \, b_3 \zeta \, + \,
   b_6 \zeta \, + \, 4 b_7 (\eta - 2 \zeta) \, - \,
   2 b_8 (\eta - 3 \zeta)}{4 b_1 + b_2 + b_3 - 2 b_6 + 4 b_7 - 4 b_8}
\end{displaymath}
lies between $\max(0, 2 \zeta - \eta)$ and $\zeta$, the circles of
intersection consist of equilibria and the higher order terms become
important, compare with~\cite{HMP19}.

\section{Equal masses in the \FPU-chain}
\label{equalmassesFPUchain}

\begin{figure}[ht]
  \centering
  \subfloat[]{\includegraphics[width=0.499\textwidth,
  height=4.99cm]{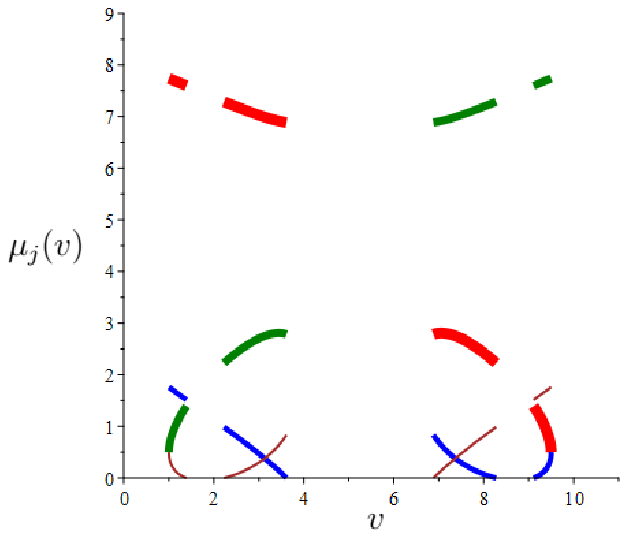}\label{figure1-a}}
  \hfill
  \subfloat[]{\includegraphics[width=0.499\textwidth,
  height=4.99cm]{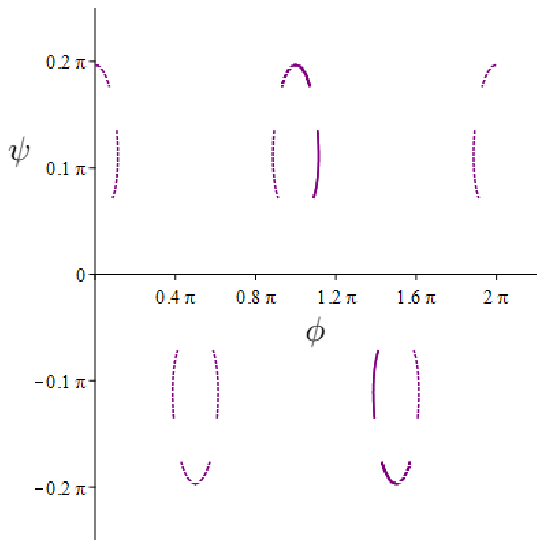}\label{figure1-b}}
  \hfill
\caption{
(a).~All the branches of the fibre of possible inverse mass
configurations of~$\mu_j(v)$, for $j = 1, 2, 3, 4$ with colors
red, green, blue, brown and thicknesses 8, 6, 4, 2, respectively.
(b).~The fibre is contained in an ellipsoid, leading to the shown
distribution of inverse masses in spherical co-ordinates with
horizontally the azimuth~$\phi$ and vertically the
inclination~$\psi$.
The continuous curves correspond to the branches of the fibre in
fig.~(a) and the dotted curves correspond to the translates of
the branches under the dihedral group~$D_4$.
All these curves on the ellipsoid show that the fibre for the
$1{:}2{:}4$~resonance consists of $12$~open curves, compare
with~\cite{BV17}.
}
\end{figure}

The parametrisation~\eqref{MassesFPU} of the inverse masses
$\mu_j$, $j = 1, \ldots, 4$ by $v \neq \frac{21}{4}$ yields
positive values $\mu_j(v) > 0$, $j = 1, \ldots, 4$ for
\begin{displaymath}
   v \;\; \in \;\; \clopin{1}{r_1} \; \cup \; \opin{r_2}{s_1}
   \; \cup \; \opin{r_3}{s_2} \; \cup \; \opclin{s_3}{\frac{19}{2}}
   \enspace ,
\end{displaymath}
see fig.~\ref{figure1-a}.
Also we can use the transforms under the group~$D_4$ (as is done
for the $1{:}2{:}3$~resonance in~\cite{BV17}),
\begin{displaymath}
   \xi_1 \; = \; \frac{- \mu_1 + \mu_2 - \mu_3 + \mu_4}{2}
   \enspace , \quad
   \xi_2 \; = \; \frac{\mu_4 - \mu_2}{\sqrt{2}}
   \enspace , \quad
   \xi_3 \; = \; \frac{\mu_3 - \mu_1}{\sqrt{2}}
   \enspace,
\end{displaymath}
where using the spherical coordinates $(\rho, \psi, \phi)$
determined by
\begin{displaymath}
   \xi_1 \; = \; \rho\sin{\psi}
   \enspace , \quad
   \xi_2 \; = \; \frac{\rho}{\sqrt{2}} \cos{\psi} \cos{\phi}
   \enspace , \quad
   \xi_3 \; = \; \frac{\rho}{\sqrt{2}} \cos{\psi} \sin{\phi}
\end{displaymath}
the distribution of inverse masses is plotted in fig.~\ref{figure1-b}.
Here as in fig.~\ref{figure1-a} the excluded boundary points are given
by
\begin{eqnarray*}
   r_1 = 1.3932195 & \quad & s_1 = 3.6182827  \\
   r_2 = 2.2250632 & \quad & s_2 = 8.2749368  \\
   r_3 = 6.8817173 & \quad & s_3 = 9.1067805  \enspace .
\end{eqnarray*}
At $v = 1$ we have $\mu_2 = \mu_4$ and at
$v = \frac{19}{2}$ we have $\mu_1 = \mu_3$.
These cases are extended cases of classic and alternating masses
cases as considered in the literature, see for instance~\cite{BV19}.
When relabeling the masses the two cases are mapped into each other
and so we may concentrate on $v = 1$.
Then the solution of the system~\eqref{MassesFPU} is
\begin{equation}
\label{equalmasses}
   (\mu_1, \mu_2, \mu_3, \mu_4) \;\; = \;\;
   (\frac{19}{4}+\frac{\sqrt{143}}{4}, \frac{1}{2},
   \frac{19}{4}-\frac{\sqrt{143}}{4}, \frac{1}{2})
\end{equation}
while $K$ and $L$ in eq.~\eqref{transformationFPU} are given by
\begin{displaymath}
   K \;\; = \;\; A_4^{-\frac{1}{2}} U \Omega^{\frac{1}{4}}
   \;\; = \;\; \left(
   \begin{array}{cccc}
      0 & \frac{\sqrt{13} - \sqrt{11}}{4 \sqrt{3}} &
      \frac{\sqrt{13} + \sqrt{11}}{4 \sqrt{6}} &
      \frac{19 - \sqrt{143}}{8 \sqrt{218}}  \\[5pt]
      1 & \frac{\sqrt{11}}{4 \sqrt{3}} &
      - \frac{\sqrt{13}}{4 \sqrt{6}} &
      \frac{\sqrt{218}}{16} \\[5pt]
      0 & - \frac{\sqrt{13} + \sqrt{11}}{4 \sqrt{3}} &
      \frac{\sqrt{13} - \sqrt{11}}{4 \sqrt{6}} &
      \frac{19 + \sqrt{143}}{8 \sqrt{218}} \\[5pt]
      -1 & \frac{\sqrt{11}}{4 \sqrt{3}} &
      - \frac{\sqrt{13}}{4 \sqrt{6}} &
     \frac{\sqrt{218}}{16}
   \end{array}
   \right)
\end{displaymath}
and
\begin{displaymath}
   L \;\; = \;\; A_4^{\frac{1}{2}} U \Omega^{-\frac{1}{4}}
   \;\; = \;\; \left(
   \begin{array}{cccc}
      0 & \frac{4 \sqrt{13} - 3 \sqrt{11}}{16 \sqrt{3}} &
      \frac{15 \sqrt{13} + 16 \sqrt{11}}{32 \sqrt{6}} &
      \frac{\sqrt{218}}{32} \\[5pt]
      \frac{1}{2} & \frac{\sqrt{11}}{16 \sqrt{3}} &
      - \frac{\sqrt{13}}{32 \sqrt{6}} &
      \frac{\sqrt{218}}{32} \\[5pt]
      0 & - \frac{4 \sqrt{13} + 3 \sqrt{11}}{16 \sqrt{3}} &
      \frac{15 \sqrt{13} - 16 \sqrt{11}}{32 \sqrt{6}} &
      \frac{\sqrt{218}}{32} \\[5pt]
      - \frac{1}{2} & \frac{\sqrt{11}}{16 \sqrt{3}} &
      - \frac{\sqrt{13}}{32 \sqrt{6}} &
      \frac{\sqrt{218}}{32}
   \end{array}
   \right)
   \enspace .
\end{displaymath}
We perturb the masses by unfolding the values in
eq.~\eqref{equalmasses} from $\mu$ to $\mu + \nu$, $\nu \in \R^4$;
in general this yields a non-diagonal matrix~$\Lambda$ if we use
the same transformation.
However, using adapted $K = K(\nu)$ and $L = L(\nu)$ resulting from
$A_4 = A_4(\nu)
 = \diag(\mu_1 + \nu_1, \mu_2 + \nu_2, \mu_3 + \nu_3, \mu_4 + \nu_4)$
and corresponding diagonalising matrix $U = U(\nu)$ again yields
$\Lambda = \diag(\gamma_1, \gamma_2, \gamma_3, 0)$ and
$\Omega = \diag(\gamma_1, \gamma_2, \gamma_3, 1)$ with now
\begin{displaymath}
   \gamma_j \;\; = \;\; (2^{j-1} + \lambda_j)^2
   \enspace , \quad j = 1, 2, 3
\end{displaymath}
where $\lambda = \lambda(\nu)$ detunes the $1{:}2{:}4$~resonance.
For $\nu = 0$ also $\lambda = 0$; the detunings $\lambda$ are
$O(\varepsilon)$ if the normalised $H_3$ does not vanish, if it
does, we have $\lambda = O(\varepsilon^2)$.
We start with neglecting the changes by detuning  in the higher
order terms, which are transformed to
\begin{eqnarray}
   H_3 & = & - \frac{\alpha \, x_1}{12}
   \left( 13 x_2 + \sqrt{286} x_3 \right)
   \left( 11 x_2 - \sqrt{286} x_3 \right)
\label{TransformedCubicHamiltonianFPU}\\
   H_4 & = & \frac{\beta}{144} \left(
   9 x_1^4 \, + \, 108 x_1^2 x_2^2 \, + \, 216 x_1^2 x_3^2
   \, + \, \frac{287}{4} x_2^4 \, - \, \sqrt{286} x_2^3 x_3
   \right. \nonumber\\
   & & \left . \qquad \, + \, 3 x_2^2 x_3^2 \, + \,
   2 \sqrt{286} x_2 x_3^3 \, + \, 287 x_3^4 \right)
\label{TransformedQuarticHamiltonianFPU}
   \enspace .
\end{eqnarray}
As already discussed, the normal form terms~$\cH_3$ of cubic order
vanish.
This in particular leads to $\lambda = O(\varepsilon^2)$.
The detuning terms are of the same order~$\varepsilon^2$ as~$\cH_4$.

\begin{Remark}
From fig.~\ref{figure1-a} we see that we also have equal masses
$\mu_3 = \mu_4$ at two values $v = 3.1305722$ and $v = 7.3694278$.
However, here the cubic normal form does not vanish.
The reason is that $\mu_3 = \mu_4$ does not define a symmetry
like the $\Z_2$--symmetry~\eqref{Z2symmetry}.
We therefore defer the discussion of these cases to the general
discussion of non-vanishing cubic normal forms in a subsequent paper.
\end{Remark}

\noindent
For the $\alpha$--chain ($\beta = 0$) at $v = 1$ we need two
normalising transformations to get the truncated normal form
$\cH^{4, \lambda} = \cH_2^{\lambda} + \cH_4$ with
$\lambda = \lambda(\nu)$ in the Hamiltonian~\eqref{Hamlam} and
\begin{eqnarray}
   b_1 & = & 0 \nonumber\\
   b_2 & = & - \frac{4147}{8640} \:\! \alpha^2 \nonumber\\
   b_3 & = & - \frac{17875}{9072} \:\! \alpha^2 \nonumber\\
   b_4 & = & - \frac{3\sqrt{286}}{280} \:\! \alpha^2 \nonumber\\
   b_5 & = & 0 \label{b1tob8-alpha}\\
   b_6 & = & \frac{15053}{7560} \:\! \alpha^2 \nonumber\\
   b_7 & = & - \frac{403}{810} \:\! \alpha^2 \nonumber\\
   b_8 & = & - \frac{209}{756} \:\! \alpha^2 \nonumber
\end{eqnarray}
in the quartic part~\eqref{Hamred4}.
For the general $\beta$--chain ($\alpha = 0$) where we do not need
to restrict to $v = 1$ to have the cubic terms vanish we get the
truncated normal form $\cH^{4, \lambda} = \cH_2^{\lambda} + \cH_4$
with $\lambda = \lambda(\nu)$ in the Hamiltonian~\eqref{Hamlam} and
\begin{eqnarray}
   b_1 & = & - \frac{3 (2v - 19)^2}{2^8 (4v - 21)^4 n_1^4}
   \left( b_{12} \Gamma^{\frac{3}{2}} \, + \,
   b_{11} \Gamma^{\frac{1}{2}} \, + \, b_{10} \right) \:\! \beta
\nonumber\\[5pt]
   b_2 & = & - \frac{3 (2v - 13)^2}{2^{10} (4v - 21)^4 n_2^4}
   \left( b_{22} \Gamma^{\frac{3}{2}} \, + \,
   b_{21} \Gamma^{\frac{1}{2}} \, + \, b_{20} \right) \:\! \beta
\nonumber\\[5pt]
   b_3 & = &  - \frac{3 (2v + 11)^2}{2^2 (4v - 21)^4 n_3^4}
  \left( b_{32} \Gamma^{\frac{3}{2}} \, + \,
   b_{31} \Gamma^{\frac{1}{2}} \, + \, b_{30} \right) \:\! \beta
\nonumber\\[5pt]
   b_4 & = & \frac{3 \Delta}{\sqrt{2}^{13} (4v - 21)^3 n_1^2 n_2 n_3}
   \left( b_{42}\Gamma^{\frac{3}{2}} \, + \,
   b_{41} \Gamma^{\frac{1}{2}} \, + \, b_{40} \right) \:\! \beta
\nonumber\\[5pt]
   b_5 & = & 0 \label{b1tob8-beta} \\[5pt]
   b_6 & = & - \frac{3 (2v + 11)(2v - 13)}{2^5 (4v - 21)^4 n_2^2 n_3^2}
   \left( b_{62} \Gamma^{\frac{3}{2}} \, + \,
   b_{61} \Gamma^{\frac{1}{2}} \, + \, b_{60} \right) \:\! \beta
\nonumber\\[5pt]
   b_7 & = & - \frac{3 (2v + 11)(2v - 19)}{2^4 (4v - 21)^4 n_1^2 n_3^2}
  \left( b_{72} \Gamma^{\frac{3}{2}} \, + \,
   b_{71} \Gamma^{\frac{1}{2}} \, + \, b_{70} \right) \:\! \beta
\nonumber\\[5pt]
   b_8 & = & - \frac{3 (2v - 13)(2v - 19)}{2^7 (4v - 21)^4 n_1^2 n_2^2}
  \left( b_{82} \Gamma^{\frac{3}{2}} \, + \,
   b_{81} \Gamma^{\frac{1}{2}} \, + \, b_{80} \right) \:\! \beta
\nonumber
\end{eqnarray}
in the quartic part~\eqref{Hamred4} where
\begin{eqnarray*}
   n_1 & = & \left( \frac{45 (2 v - 19) [ (6 v^2 - 57 v + 107)
   \sqrt{\Gamma} \, + \, 6 v^3 - 69 v^2 + 229 v - 256 ]}{2 (- v +
   \sqrt{\Gamma})(4 v - 21)} \right)^{\frac{1}{2}} \\
   n_2 & = & \left( - \frac{9 (2 v - 13) [ (6 v^2 - 39 v - 52)
   \sqrt{\Gamma} \, + \, 6 v^3 - 87 v^2 + 388 v - 256]}{2 ( - v +
   \sqrt{\Gamma})(4 v - 21)} \right)^{\frac{1}{2}} \\
   n_3 & = & \left( \frac{90 (2 v + 11) [ (6 v^2 + 33 v - 328)
   \sqrt{\Gamma} \, + \, 6 v^3 - 159 v^2 + 664 v - 256]}{( - v +
   \sqrt{\Gamma})(4 v - 21)} \right)^{\frac{1}{2}}
\end{eqnarray*}
are defined as in~\cite{BV17} (together with an $n_4$ that we do not
need here) and
\begin{eqnarray*}
   b_{12} & = & 2 (4 v - 21)^2
   (48 v^5 - 584 v^4 + 3932 v^3 - 85074 v^2 + 746170 v - 1877017) \\
   b_{11} & = & 8 (4 v - 21)
   (72 v^8 - 1948 v^7 + 18782 v^6 - 55941 v^5 - 331278 v^4 + 3089076 v^3 \\
   & & \mbox{} \qquad \qquad - 6496428 v^2 - 10260437 v + 40373227) \\
   b_{10} & = & - (48 v^6 - 1512 v^5 + 22576 v^4 - 196266 v^3 +
   898339 v^2 - 1676472 v + 368062) \\
   & & \mbox{} \!\!\!\! \times \;
   (28 v^4 - 900 v^3 + 6419 v^2 - 12950 v - 697) \\
   b_{22} & = & 4 (4 v - 21)^2
   (24 v^5 - 412 v^4 - 422 v^3 + 43143 v^2 - 298372 v + 676534) \\
   b_{21} & = & 4 (4 v - 21)
   (144 v^8 - 5000 v^7 + 66532 v^6 - 368922 v^5 - 143706 v^4 + 12250443v^3 \\
   & & \mbox{} \qquad \qquad - 61549224 v^2 + 133760048 v - 122081152) \\
   b_{20} & = & - (48 v^6 - 1512 v^5 + 19504 v^4 - 131754 v^3 +
   436867 v^2 - 387240 v - 682400) \\
   & & \mbox{} \!\!\!\! \times \;
   (28 v^4 - 828 v^3 + 5999 v^2 - 15848 v + 8048) \\
   b_{32} & = & 4 (4 v - 21)^2
   (24 v^5 - 892 v^4 + 18826 v^3 - 29937 v^2 - 844720 v + 3115234) \\
   b_{31} & = & 4 (4 v - 21)
   (144 v^8 - 9416 v^7 + 245764 v^6 - 3135882 v^5 + 16325934 v^4 +
   67107387 v^3 \\
   & & \mbox{} \qquad \qquad - 1318986636 v^2 + 6135153536 v - 9655555456) \\
   b_{30} & = & - (48 v^6 - 1512 v^5 + 18736 v^4 - 115626 v^3 -
   668141 v^2 + 10326288 v - 28221968) \\
   & & \mbox{} \!\!\!\! \times \;
   (28 v^4 - 540 v^3 - 721 v^2 + 33040 v - 96832) \\
   b_{42} & = & (4 v - 21)^2
   (48 v^4 - 920 v^3 + 6676 v^2 - 21436 v + 24965) \\
   b_{41} & = & 2 (v - 1) (4 v - 21)
   (144 v^6 - 5336 v^5 + 76684 v^4 - 547544 v^3 + 2027809 v^2 \\
   & & \mbox{} \qquad \qquad \qquad \qquad
   - 3622625 v + 2366158) \\
   b_{40} & = & - (v - 1) (v - 4) (v - 16) (2 v - 19)
   (12 v^2 - 126 v + 277) \\
   & & \mbox{} \!\!\!\! \times \;
   (28 v^3 - 512 v^2 + 2667 v - 3713) \\
   b_{62} & = & 4 (2 v - 19) (4 v - 21)^2
   (12 v^4 - 212 v^3 + 1435 v^2 - 6682 v + 16004) \\
   b_{61} & = & 4 (2 v - 19) (4 v - 21)
   (72 v^7 - 2920 v^6 + 43422 v^5 - 282828 v^4 + 593619 v^3 \\
   & & \mbox{} \qquad \qquad \qquad \qquad
   + 1700190 v^2 - 8425848 v + 6257248) \\
   b_{60} & = & - (48 v^6 - 1512 v^5 + 14512 v^4 - 26922 v^3 -
   256469 v^2 + 1113924 v - 830576) \\
   & & \mbox{} \!\!\!\! \times \;
   (28 v^4 - 684 v^3 + 5231 v^2 - 11276 v - 6304) \\
   b_{72} & = & (2 v - 13) (4 v - 21)^2
   (48 v^4 - 872 v^3 + 7924 v^2 - 44368 v + 101273) \\
   b_{71} & = & 2 (2 v - 13) (4 v - 21)
   (144 v^7 - 5720 v^6 + 82884 v^5 - 525336 v^4 + 750405 v^3 \\
   & & \mbox{} \qquad \qquad \qquad \qquad
   + 8645652 v^2 - 47236533 v + 73402004) \\
   b_{70} & = & - (48 v^6 - 1512 v^5 + 13456 v^4 - 4746 v^3 -
   542351 v^2 + 2893233 v - 4699028) \\
   & & \mbox{} \!\!\!\! \times \;
   (28 v^4 - 720 v^3 + 6899 v^2 - 29105 v + 45848) \\
   b_{82} & = & (2 v + 11) (4 v - 21)^2
   (48 v^4 - 968 v^3 + 6580 v^2 - 16648 v + 12161) \\
   b_{81} & = & 2 (2 v + 11) (4 v - 21)
   (144 v^7 - 5240 v^6 + 80004 v^5 - 658896 v^4 + 3142173v^3 \\
   & & \mbox{} \qquad \qquad \qquad \qquad
   - 8672760 v^2 + 12977859 v - 8267824) \\
   b_{80} & = & - (48 v^6 - 1512 v^5 + 20752 v^4 - 157962v^3 +
   687961 v^2 - 1579011 v + 1497904) \\
   & & \mbox{} \!\!\!\! \times \;
   (28 v^4 - 864 v^3 + 6371 v^2 - 17261 v + 16316)
  \enspace .
\end{eqnarray*}
Interestingly, at $v = 1$ we have $b_4 = 0$.
Since $\Delta$ is a factor of~$b_4$ we also have $b_4 = 0$ at
$v = \frac{19}{2}$ where we furthermore have $b_1 = b_7 = b_8 = 0$.
Adding the $\alpha$-- and $\beta$--chain, i.e.\ working with the
full potential~\eqref{FPUpotential} we get at $v = 1$ the
coefficients
\begin{eqnarray}
   b_1 & = & \frac{3}{2^6} \:\! \beta \nonumber\\
   b_2 & = & - \frac{11 \cdot 13 \cdot 29}{2^6 3^3 5} \:\! \alpha^2
   \; + \; \frac{7 \cdot 41}{2^8 3} \:\! \beta \nonumber\\
   b_3 & = & - \frac{5^3 \cdot 11 \cdot 13}{2^4 3^4 7} \:\! \alpha^2
   \; + \; \frac{7 \cdot 41}{2^6 3} \:\! \beta \nonumber\\
   b_4 & = & - \frac{3\sqrt{2 \cdot 11 \cdot 13}}{2^3 \cdot 5 \cdot 7}
   \:\! \alpha^2 \nonumber\\
   b_5 & = & 0 \label{b1tob8}\\
   b_6 & = & \frac{15053}{2^3 \cdot 3^3 \cdot 5 \cdot 7} \:\! \alpha^2
   \; + \; \frac{1}{2^5 3} \:\! \beta \nonumber\\
   b_7 & = & - \frac{13 \cdot 31}{2 \cdot 3^4 \cdot 5} \:\! \alpha^2
   \; + \; \frac{3}{2^2} \:\! \beta \nonumber\\
   b_8 & = & - \frac{11 \cdot 19}{2^2 3^3 7} \:\! \alpha^2 \; + \;
   \frac{3}{2^3} \:\! \beta \nonumber
\end{eqnarray}
in the quartic part~\eqref{Hamred4}.
Note that both $\alpha^2$ and~$\beta$ are of order~$O(\varepsilon^2)$.

\section{Dynamics in one degree of freedom}
\label{dynamics1dof}
Here we study the dynamics of a general Hamitonian
\begin{equation}
\label{genHam}
   \cH \;\; = \;\; \sigma_5 \; + \; p_{\eta, \zeta}^{\lambda} \tau_2
   \; + \; q \frac{\tau_2^2}{2}
\end{equation}
on the reduced phase space, a surface given by eq.~\eqref{redphaseineq}.
We defer to the end of this section the discussion of the dependence
of the parameters $p_{\eta, \zeta}^{\lambda}$ and~$q$ via
eq.~\eqref{redHam} on the detuning $\lambda = \lambda(\nu)$, the
external parameters $\alpha$, $\beta$ and $\mu_j$, $j = 1, \ldots, 4$
of the \FPU-chain and the distinguished parameters $\eta$ and~$\zeta$.
In particular, we assume that the coefficient $b_4$ does not vanish
in the truncated normal form~\eqref{redHam}.
The orbits of the equations of motion defined by the
Hamiltonian~\eqref{genHam} coincide with the intersections of the
surface~\eqref{redphaseineq} with the energy level sets $\{ \cH = h \}$.

\begin{figure}[htb]
\begin{center}
\begin{picture}(471,160)
   \put(0,0){\includegraphics[height=5cm,keepaspectratio]{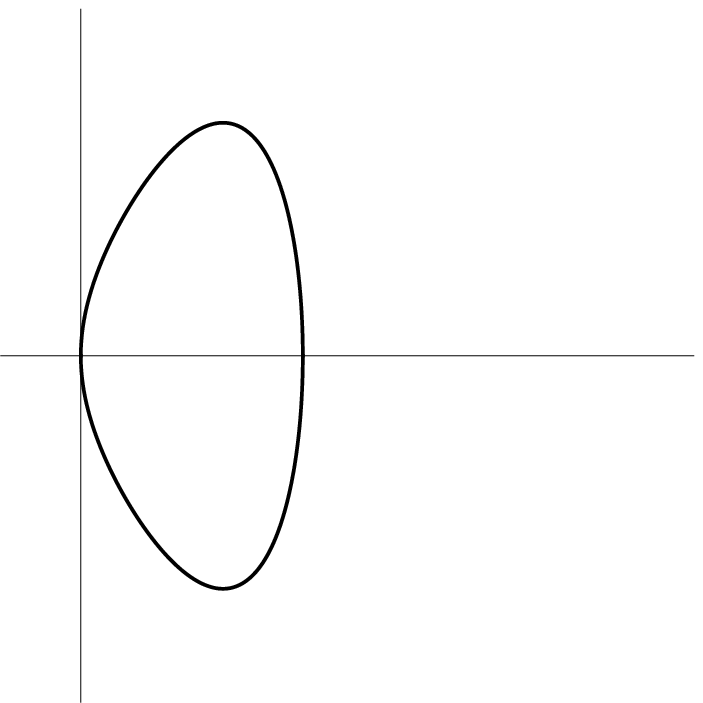}}
   \put(165,0){\includegraphics[height=5cm,keepaspectratio]{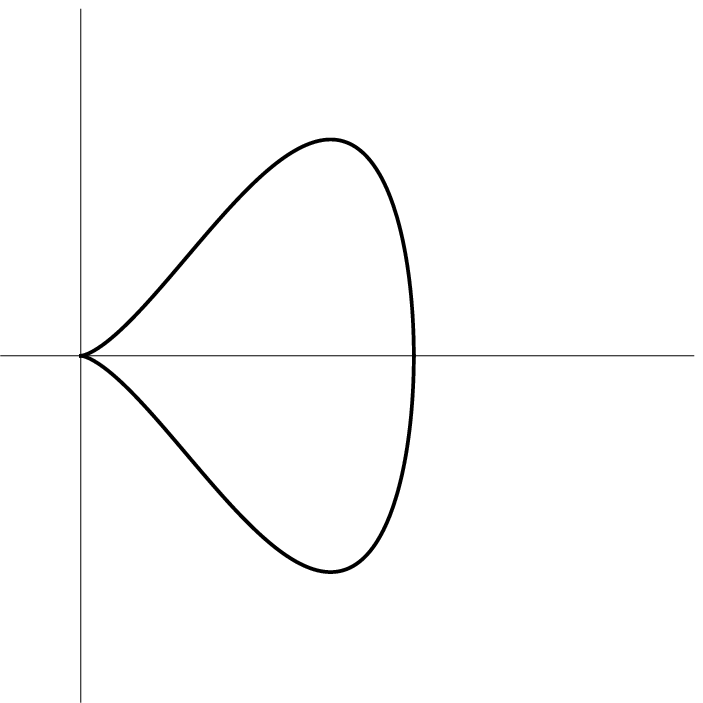}}
   \put(330,0){\includegraphics[height=5cm,keepaspectratio]{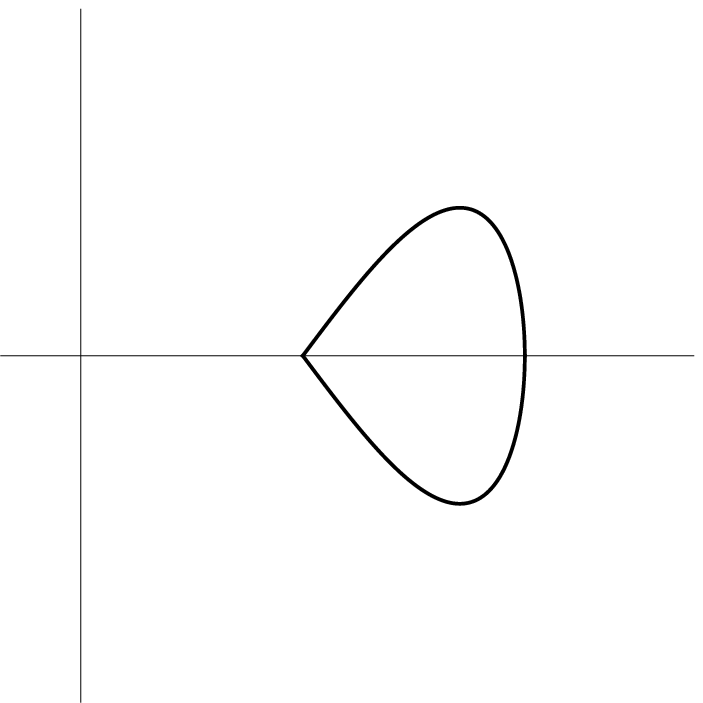}}
\end{picture}
\end{center}
\caption{
The possible forms of the reduced phase space~\eqref{redphaseineq}
in the intersection with $\{ \sigma_6 = 0 \}$ given by the
curves~\eqref{intercurve}.
To get back the surface of revolution~\eqref{redphaseineq} one has
to rotate the curve(s) about the horizontal axis.
}\label{figure2}
\end{figure}

The singular points of the reduced phase space~\eqref{redphaseineq}
are always equilibria, these occur where $\tau_2 = 2 \zeta - \eta \geq 0$.
Indeed, for $2 \zeta = \eta$ this value is~$0$ and the surface of
revolution has a cuspoidal singular point, while for $2 \zeta > \eta$
this value is positive and the singular point is conical, compare
with fig.~\ref{figure2}.
At $\zeta = \eta$ the reduced phase space~\eqref{redphaseineq}
shrinks down to a single point, from which the normal--$2$--mode is
reconstructed, and also when $\zeta = 0$ the reduced phase space is
a singular point, corresponding to the normal--$1$--mode
$\tau_1 = 2 \eta$.
The normal--$3$--mode $\tau_3 = \half \eta$ is reduced to the
cuspoidal singular point $\tau_2 = 0$ at the origin
of~\eqref{redphaseineq} in case $2 \zeta = \eta$.
As all three normal modes yield singular points in $1$~dof, the
reconstruction of~\eqref{genHam} to $3$~dof has three families of
periodic orbits in the co-ordinate planes $(x_j, y_j)$, i.e.\ at
the normal modes.

For regular equilibria in $1$~dof the energy level set
$\{ \cH = h \}$ has to be tangent to the reduced phase
space~\eqref{redphaseineq}.
The latter is a surface of revolution and the former is a cylinder
in the $\sigma_6$--direction --- on the basis of the intersection
(a parabola)
\begin{equation}
\label{intersect}
   \{ \cH = h \} \; \cap \; \{ \sigma_6 = 0 \} \;\; : \;\;
   \sigma_5 \; = \; h - p_{\eta, \zeta}^{\lambda} \tau_2
   - q \frac{\tau_2^2}{2}
   \enspace .
\end{equation}
Hence, equilibria $(\tau_2^*, \sigma_5^*, \sigma_6^*)$ necessarily
satisfy $\sigma_6^* = 0$.
The value $h$ of~$\cH$ can always be adjusted to let the reduced
phase space~\eqref{redphaseineq} and the level sets of the
Hamiltonian~\eqref{genHam} intersect.
The two curves that form the intersection of the
surface~\eqref{redphaseineq} with $\{ \sigma_6 = 0 \}$ are given by
\begin{equation}
\label{intercurve}
   \sigma_5 \; = \;
   \pm 4 (\eta - 2 \zeta + \tau_2) \sqrt{\tau_2 (\zeta - \tau_2)}
   \enspace , \quad
   \max(0, 2 \zeta - \eta) \leq \tau_2 \leq \zeta
\end{equation}
and have derivatives
\begin{equation}
\label{derivcurve}
   \frac{\D \sigma_5}{\D \tau_2} \;\; = \;\;
   \pm 2 \left[ 2 \sqrt{\tau_2 (\zeta - \tau_2)} \; + \;
   \frac{(\eta - 2 \zeta + \tau_2)(\zeta - 2 \tau_2)}
   {\sqrt{\tau_2 (\zeta - \tau_2)}} \right]
   \enspace .
\end{equation}
We may concentrate on the derivatives of~$\sigma_5$ with respect
to~$\tau_2$ in eq.~\eqref{intersect} and only need to solve the
equation
\begin{equation}
\label{derivequal}
   \mp \sqrt{\tau_2 (\zeta - \tau_2)}
   (p_{\eta, \zeta}^{\lambda} + q \tau_2)
   \;\; = \;\; 4 \tau_2 (\zeta - \tau_2) \; + \;
   2 (\eta - 2 \zeta + \tau_2)(\zeta - 2 \tau_2)
\end{equation}
for $\tau_2^*$ between $\max(0, 2 \zeta - \eta)$ and~$\zeta$ to
obtain the values $(\tau_2^*, \sigma_5^*, 0)$ of equilibria, with
$\sigma_5^*$ given by eq.~\eqref{intercurve}.
The regular equilibria undergo a centre-saddle bifurcation where the
curves given by \eqref{intersect} and~\eqref{intercurve} not only
touch but also have coinciding curvature, resulting in the equation
\begin{equation}
\label{curvaequal}
   \mp q \sqrt{\tau_2^*}^3 \sqrt{\zeta - \tau_2^*} \;\; = \;\;
   4 \tau_2^* (\zeta - 2 \tau_2^*) \; - \;
   4 \tau_2^* (\eta - 2 \zeta + \tau_2^*) \; - \;
   3 (\eta - 2 \zeta + \tau_2^*) (\zeta - 2 \tau_2^*)
\end{equation}
on $\eta$ and~$\zeta$.
In case the condition~\eqref{derivequal} holds at the singular point
$\tau_2^* = 2 \zeta - \eta > 0$ this equilibrium undergoes a
Hamiltonian flip bifurcation (which reconstructs to a period
doubling bifurcation in $2$~dof and to a frequency-halving
bifurcation in $3$~dof).
Where this happens at $\tau_2^* = 0$ the energy level
curve~\eqref{intersect} passes horizontally through the singular
point reduced from the normal--$3$--mode, resulting in an unstable
manifold and thus revealing the normal--$3$--mode to be unstable.
Next to $\eta = 2 \zeta$ (to make $\tau_2 = 0$ a singular point
of the reduced phase space) this requires
$p_{\eta, \zeta}^{\lambda} = 0$, i.e.\ for the normal
form~\eqref{redHam} that
\begin{displaymath}
   (b_3 \, - \, b_6 \, - \, 2 b_8) \zeta \;\; = \;\;
   2 \lambda_1 \; + \; \lambda_2 \; - \; \lambda_3
   \enspace .
\end{displaymath}
It is helpful to illustrate the dynamics using the
contour~\eqref{intercurve}, see fig.~\ref{figure2}.
Indeed, from these sections one can easily construct the surface
of revolution~\eqref{redphaseineq}, the intersections of which with
the parabolic cylinders $\{ \cH = h \}$ yield the orbits of the
reduced dynamics.
The parabolic cylinders are determined by the
parabolas~\eqref{intersect} whence the relative positions of the
contour~\eqref{intercurve} and these parabolas allow to get a full
picture of the reduced flow on the surface~\eqref{redphaseineq}.
For $q > 0$ the parabolas~\eqref{intersect} are `upside-down', with a
maximum at
\begin{displaymath}
   \tau_2 \;\; = \;\; - \frac{p_{\eta, \zeta}^{\lambda}}{q}
   \enspace .
\end{displaymath}
Hence, regular equilibria $(\tau_2^*, \sigma_5^*, 0)$ on the lower
arc of the contour~\eqref{intercurve} are elliptic (sometimes the
miniml value $h$ of~$\cH$ is taken at the singular point
$\tau_2^* = 2 \zeta - \eta$) while on the upper arc we may
have both hyperbolic and elliptic equilibria, depending on the
centre-saddle and Hamiltonian flip bifurcations (in particular also
the maximal value $h$ of~$\cH$ is sometimes taken at the singular
point $\tau_2^* = 2 \zeta - \eta$).

\begin{figure}
\begin{center}
\begin{picture}(430,95)
   \put(0,0){\includegraphics[height=3.25cm,keepaspectratio]{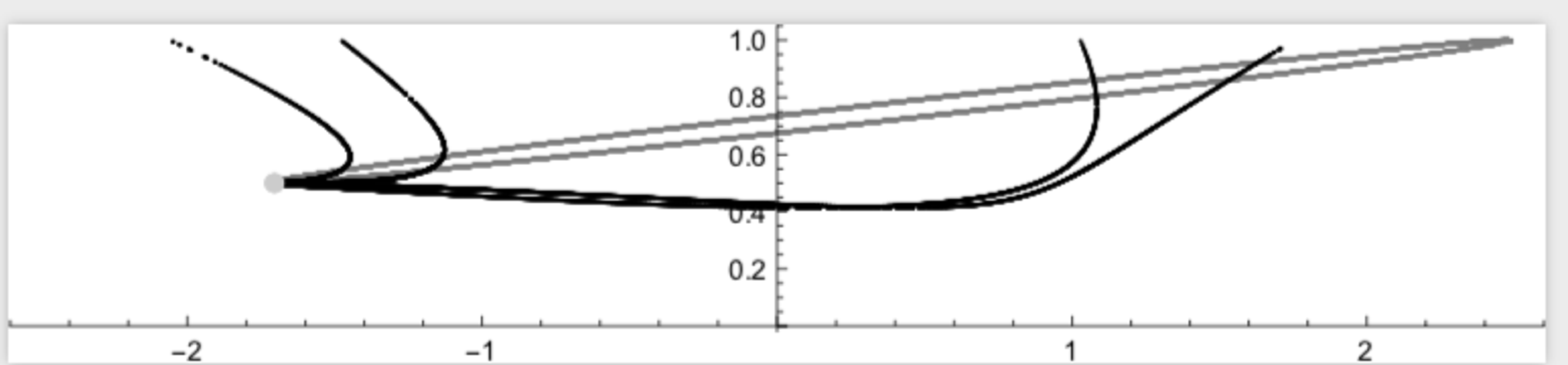}}
\end{picture}
\end{center}
\caption{
The bifurcation diagram of the dynamics defined by the normal form
of the Hamiltonian function~\eqref{FPUHam} with $v = 1$ in $1$~dof
for $(\alpha, \beta) = (1, 0)$ --- the $\alpha$--chain --- where we
put $2 \lambda_1 + \lambda_2 - \lambda_3$ on the horizontal axis,
while we normed $\eta = 1$ and put $\zeta \in [0, 1]$ on the vertical
axis.
The black lines are centre-saddle bifurcations and the dark grey
lines form a circle of Hamiltonian flip bifurcations.
The light grey point from which all these lines emerge is where
the normal--$3$--mode is momentarily unstable.
}\label{figure3}
\end{figure}

The Hamiltonian flip bifurcation occurs where the
parabola~\eqref{intersect} enters the singular point
$\tau_2^* = 2 \zeta - \eta > 0$ with the same slope as the upper
or lower arc of the contour~\eqref{intercurve}.
Between these two curves in the $(\zeta, \eta)$--plane the singular
equilibrium $\tau_2^* = 2 \zeta - \eta > 0$ is unstable.
Together with the curves~\eqref{curvaequal} of centre-saddle
bifurcations these yield the bifurcation diagram shown in
fig.~\ref{figure3}.
For a centre-saddle bifurcation the parabola~\eqref{intersect} and
the contour~\eqref{intercurve} have to pass each other
non-transversely --- the equilibrium where the bifurcation takes
place requires that the two curves have to touch each other.

To actually compute the bifurcation diagram, the precise dependence
of $p_{\eta, \zeta}^{\lambda}$ and~$q$ on the external parameters
becomes important.
For instance, for the normal form~\eqref{redHam} we have
\begin{align}
   p_{\eta, \zeta}^{\lambda} & \;\; = \;\; \frac{
   2 \lambda_1 + \lambda_2 - \lambda_3 \, + \,
   4 b_1 (\eta - 2 \zeta) \, - \, b_3 \zeta
   \, + \, b_6 \zeta \, + \, 4 b_7 (\eta - 2 \zeta)
   \, - \, 2 b_8 (\eta - 3 \zeta)}{b_4} \\
\intertext{and}
   q & \;\; = \;\;
   \frac{4 b_1 + b_2 + b_3 - 2 b_6 + 4 b_7 - 4 b_8}{b_4}
   \enspace .
\end{align}
Note that the detuning $\lambda \in \R^3$ only enters through the
combination $2 \lambda_1 + \lambda_2 - \lambda_3$.
In particular, the $\alpha$--chain with $v = 1$ has $b_1, \ldots, b_8$
given by~\eqref{b1tob8-alpha}, whence
\begin{displaymath}
   p_{\eta, \zeta}^{\lambda} \; = \;
   - \frac{- 45360 (2 \lambda_1 + \lambda_2 - \lambda_3) \, + \,
   (65192 \eta - 284997 \zeta) \alpha^2}{486 \sqrt{286} \:\, \alpha^2}
   \quad \mbox{and} \quad
   q \; = \; \frac{1327579}{1944 \sqrt{286}}
   \enspace .
\end{displaymath}
The bifurcation diagram depicted in fig.~\ref{figure3} is for
the $\alpha$--chain with $v = 1$ for $\alpha = 1$.
Using~\eqref{b1tob8-beta} one can compute $p_{\eta, \zeta}^{\lambda}$
and~$q$ also for the general $\beta$--chain but this leads to bulky
formulas.

\section{Reconstruction of the dynamics}
\label{reconstructiondynamics}

In this section we reconstruct to higher~dof.
Reconstructing a degree of freedom amounts to replacing each
(regular) point by a circle.
The cyclic angle on this circle carries the dynamics that has
been reduced.
Where the symmetric dynamics approximates a non-symmetric system,
the former then is subject to a perturbation analysis to reveal
the structure of the latter.

\subsection{Reconstruction to $2$ degrees of freedom}
\label{reconstruction2dof}

While $\cH_2^0$ is always an integral of the normal form, the
integral~$\cK$ of~$\cH^{4, \lambda}$ is typically made deficient by
higher order normal forms (and also by a cubic normal form with
non-vanishing~$\cH_3$).
Therefore we first reconstruct the dynamics in $2$~dof
with the symmetry generated by~$\cH_2^0$ still reduced.
This consists of attaching an~$\pS$ to every regular point of the
surface of revolution~\eqref{redphaseineq}, thereby reconstructing
periodic orbits in $2$~dof from regular equilibria in $1$~dof.
From the singular point $\tau_2 = 2 \zeta - \eta > 0$ we also
reconstruct a periodic orbit, but of half the period (recall that
here the $\pS$--action generated by~$\cK$ has isotropy~$\Z_2$).
From the cuspoidal point $\tau_2 = 2 \zeta - \eta = 0$ we
reconstruct the normal--$3$--mode, while the normal--$1$ and
$2$--modes are reconstructed from the single-point versions
of~\eqref{redphaseineq} with $\zeta = 0$ and $\eta = \zeta$,
respectively --- where this happens simultaneously we have the
$1{:}2{:}4$~resonant equilibrium, which in $2$~dof constitutes
$\{ \eta = 0 \}$.

The higher order terms in the normal form yield a slow perturbation
of the semi-slow dynamics reconstructed from $1$~dof.
In $2$~dof a smooth family of invariant tori is not structurally
stable with respect to such integrability-breaking perturbations.
However, the periodic orbits constituting the bifurcation diagram
of fig.~\ref{figure3} do persist under such perturbations, whence,
up to Cantorising families of invariant tori by Diophantine
conditions, the integrable dynamics in $1$~dof also governs the
perturbed dynamics in $2$~dof.

\subsection{Reconstruction to $3$ degrees of freedom}
\label{reconstruction3dof}

When reconstructing the fast motion along the flow of~$\X{H_2^0}$,
the normal modes change from being equilibria in $2$~dof to
periodic orbits in $3$~dof and only the $1{:}2{:}4$~resonant
equilibrium remains an equilibrium.
By the same token, the periodic orbits in $2$~dof reconstruct to
invariant $2$--tori in $3$~dof and maximal tori reconstruct to
maximal tori (of now dimension~$3$, with one fast and two semi-slow
frequencies).

As any normal form $\cH$ of~$H$ is by definition equivariant with
respect to the flow of~$\X{H_2^0}$, the symmetry-breaking terms lead
to an exponentially small perturbation.
Resonances among the two-timescale-frequencies of maximal tori need
a rather high order $k \in \Z^3$, but the resonant $3$--tori still
break up and lead to (rather small) gaps.
{\it Mutatis mutandi} for resonant invariant $2$--tori: for elliptic
and hyperbolic tori persistence follows from the theory
in~\cite{BHS96} and for the quasi-periodic centre-saddle bifurcations
and the frequency halving bifurcations persistence follows from the
theory in~\cite{HH07}.

\subsection{Reconstruction to $4$ degrees of freedom}
\label{reconstruction4dof}

Reconstructing the $\pS$-symmetry~\eqref{S1symmetry} amounts to
rotating the ring of $4$~masses with a velocity governed by
$y_4 = \Sum p_j$.
This last reconstruction to the full inhomogeneous \FPU-chain in
$4$~dof is not accompanied by a perturbation analysis as the
$\pS$--symmetry~\eqref{S1symmetry} is not only a symmetry of the
normal form, but a symmetry of the original system as well.
Invariant tori reconstruct to invariant tori of one more dimension
(which may now have one resonance), in particular the normal modes
become invariant $2$--tori and the maximal tori have dimension~$4$.
In fact, the dynamics of the inhomogeneous \FPU-chain with $4$~masses
is best understood after reduction to $3$~dof.

\section{Conclusions}
\label{conclusions}

The reduced $3$~dof inhomogeneous, spatially periodic \FPU-chain
with two equal masses studied here has an integrable normal form.
For the $1{:}2{:}4$~resonance of this chain this holds true in the
case of two opposing equal masses, but not for two adjacent equal
masses.

The symmetry of the equations of motion induced by our assumption
of two opposing equal masses triggers off the existence of the
three normal mode periodic solutions.
In the original $4$--particles system the normal modes can be
reconstructed as a periodic mixture of the particle solutions.

It is expected that breaking the symmetry induced by two equal
masses will produce interesting bifurcations.
This also gives rise to integrability-breaking phenomena of the
normal forms.

\bigskip
\noindent
{\bf Acknowledgements.}
We thank Roelof Bruggeman and Evelyne Hubert for helpful discussions.

\end{document}